\documentclass{IEEEcsmag}
\usepackage[colorlinks,urlcolor=blue,linkcolor=blue,citecolor=blue]{hyperref}
\expandafter\def\expandafter\UrlBreaks\expandafter{\UrlBreaks\do\/\do\*\do\-\do\~\do\'\do\"\do\-}
\usepackage{graphicx} %
\usepackage[caption=false, font=footnotesize]{subfig}
\usepackage{hyperref}
\usepackage{braket}
\usepackage[frozencache=true,cachedir=.]{minted}

\graphicspath{ {./images/} }

\usepackage{xcolor}

\sptitle{Article Type: Description  (see Introduction for more detail)}

\jvol{XX}
\jnum{XX}
\paper{8}
\jmonth{Month}
\jname{Publication Name}
\jtitle{Publication Title}
\pubyear{2023}

\begin{document}

\title{Quantum Computing for All: Online Courses Built Around Interactive Visual Quantum Circuit Simulator}

\author{Juha Reinikainen}
\affil{Faculty of Information Technology, University of Jyväskylä, Finland}

\author{Vlad Stirbu}
\affil{Faculty of Information Technology, University of Jyväskylä, Finland}

\author{Teiko Heinosaari}
\affil{Faculty of Information Technology, University of Jyväskylä, Finland}

\author{Vesa Lappalainen}
\affil{Faculty of Information Technology, University of Jyväskylä, Finland}

\author{Tommi Mikkonen}
\affil{Faculty of Information Technology, University of Jyväskylä, Finland}

\markboth{THEME/FEATURE/DEPARTMENT}{THEME/FEATURE/DEPARTMENT}

\begin{abstract}%
Quantum computing is a highly abstract scientific discipline, which, however, is expected to have great practical relevance in future information technology.
This forces educators to seek new methods to teach quantum computing for students with diverse backgrounds and with no prior knowledge of quantum physics. 
We have developed an online course built around an interactive quantum circuit simulator designed to enable easy creation and maintenance of course material with ranging difficulty. The immediate feedback and automatically evaluated tasks lowers the entry barrier to quantum computing for all students, regardless of their background.
\end{abstract}

\maketitle

\chapteri{Q}uantum computing is a rapidly developing field that has the potential to transform the way we approach complex computational problems. By harnessing the principles of quantum mechanics, quantum computers may be able to solve problems that are currently unsolvable or require an impractical amount of time to solve using classical computers.

Quantum computing is a highly mathematical and abstract field that requires a strong foundation in physics, computer science, and mathematics. The concepts are often difficult to grasp, especially for students who have not had prior exposure to quantum mechanics or linear algebra. Therefore, it is often seen as a purely theoretical subject, and it can be challenging to relate the concepts to real-world applications or to explain how they can be used to solve practical problems. 

Since quantum computing is becoming a practical field, it requires students to gain hands-on experience with both quantum software and hardware. However, the learning curve for quantum computing should be smoother and not require deep programming expertise from the start. Further, even if there are already some quantum computers available for online access, the computing time may be limited, especially for students who are not enrolled in specialized programs.
These reasons have motivated the University of Jyväskylä to develop an interactive quantum circuit simulator as part of the existing learning platform.

The rest of paper is structured as follows, aligning with the phases of the Design Science Research (SDR) methodology \cite{doi:10.2753/MIS0742-1222240302}. We start by introducing the problem identification and motivation, followed by the learning environment in which the solution is integrated. Then, we present the quantum simulator implementation, and describe how it is used to conduct three specific tasks within the educational process. Next, we discuss how the visual editor meets the teaching goals for the target student audience. We finish by exploring the feedback received from the first group of users that completed the course in which the system was used.

\section{MOTIVATION}
\label{sec:motivation}

Quantum computing is an interdisciplinary topic that cannot be put under a single scientific field.
For that reason, also students from various study tracks are eager to learn about quantum computing. 
However, their different background and prerequisite knowledge poses an obvious challenge.
A physics student knows about quantum mechanics, but may not be fluent in programming. 
A computer science student knows programming, but has not the needed theoretical background of quantum physics.
Further, a third case is e.g. a student in business school, who understands how technology relates to business, and would hence keen to know the basics of quantum computing. 
These student personas, their educational background and expertise, as well as their expectations are summarized in Table \ref{table:personas}.
When developing the new teaching material, the different personas have served as imaginary student profiles and helped us to take into account their diverse viewpoints.

\begin{table*}
\vspace*{4pt}
\caption{Student personas, their knowledge, expertise, and expected results from using the educational system to learn quantum computing (QC).}
\label{table:personas}
\tablefont
\begin{tabular*}{\textwidth}{@{}p{100pt}p{160pt}<{\raggedright}p{160pt}<{\raggedright}@{}}
\toprule
Student persona & 
Background & 
Expectations\\
\colrule
Physics & Knows the basics of quantum mechanics & QC is not only a paper exercise \\
Software engineer & Knows programming and development processes & Gets familiar with the theoretical background of the field \\
Business & Knows how to use technology in commercial context &  Understands the challenges of QC \\
\botrule
\end{tabular*}\vspace*{8pt}
\end{table*}

To lower the entry barrier for both the students and teaching staff, we decided to extend the learning platform TIM for teaching also the quantum computing course.
As the TIM platform is used at the university for various courses, especially programming courses, the staff and major part of the students are already familiar with it.
The TIM platform needed an extension to be able to have interactive quantum circuit exercises and examples. The interactive quantum circuit simulator is, in fact, the most important part of the online course. 
It can also be used as a stand-alone tool when demonstrating quantum algorithms or other quantum information protocols in various courses or events.

\section{LEARNING ENVIRONMENT}
\label{sec:environment}

The Interactive Material (TIM)~\cite{TIMLessIsMore2019} is a massive open online course (MOOC) learning environment.
The development of TIM began in 2014 when there was no ready-made tool suitable for programming courses that would make it easy to produce and maintain %
long interactive book-like learning materials. TIM has been an open-source application under the MIT license from the beginning\footnote{\href{https://github.com/TIM-JYU/TIM}{https://github.com/TIM-JYU/TIM}}. The development of TIM is the responsibility of the staff at the Information Technology faculty. A large part of TIM has been developed as theses in computer science education.

The basic idea of TIM is document-based. Everything done in TIM is essentially documents. Documents consist of blocks. 
A block can be one or more normal text paragraphs or an interactive element. 
An interactive element can be, for example, a multiple-choice task, a drawing task, a free text field, a programming code task, an image, a series of images, a video, a mathematically handwritten task returned as an image, an assisted LaTeX-written answer, or a new component. 
The appearance of TIM is adjusted with CSS styles, so users can be offered ready-made appearance styles or users can create their own style sheet. 
Teachers can choose the appearance from ready-made styles or create their own style for their course. 

\begin{figure}
    \centering
    \includegraphics[width=\columnwidth]{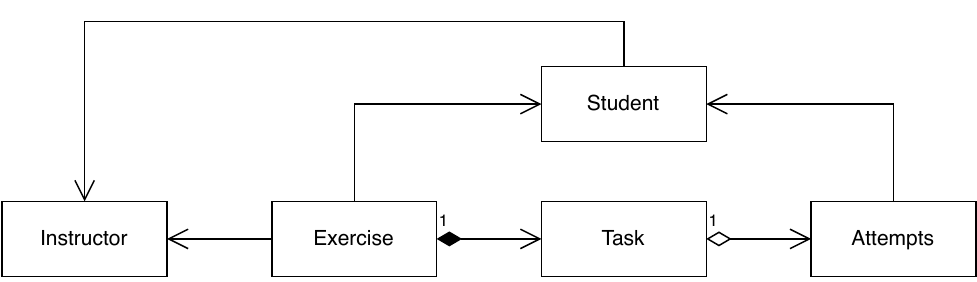}
    \caption{Information model supporting instructor-student interactions}
    \label{fig:information-model}
\end{figure}

TIM can be used to create lecture materials according to the original plan, but in the ten years TIM has developed so that it can handle all the necessary course bookkeeping from individual tasks done in TIM to the final grade. The instructor can see all the attempts the student has made on a specific task and can use it to try to figure out why the student may initially go in the wrong direction. The instructor can write feedback on the student’s answers. This way, TIM provides a lot of data for learning analytics. Students can comment on the material and the instructor can respond to the comments. In fact, this is currently the most common way to communicate with students using TIM. The information model that supports this interaction is presented in Figure~\ref{fig:information-model}.

TIM documents can be used to borrow blocks from one document to another, and thus with a single maintenance, smaller documents suitable for different needs can be made from one document, or vice versa, smaller documents can be combined into cohesive entities. Thanks to macros, parts that depend on years or even tool version numbers can be combined into one place in TIM, thus speeding up maintenance as things change. The TIM document can be translated into several different languages using the automatic translation provided by DeepL\footnote{\href{https://www.deepl.com/translator}{https://www.deepl.com/translator}}. Naturally, the human must check the result produced by the translator. If changes are made to the original document, the translation administrator will be notified and new blocks or a note of the changed section will be added to the translated document.

When using TIM, the teacher and student can do everything in one system without the need to switch from one system to another. TIM is suitable and has been used in all kinds of teaching methods from classroom teaching to MOOC courses. TIM is at its best in teaching where there are performers in many different ways. Conditional blocks allow different groups to be offered slightly different instructions if necessary.

\section{ARTIFACT DESIGN}

We implemented quantum circuit simulator in TIM platform, so that it can be directly used for developing the quantum technology curriculum.

At the start of the project, TIM already had all the essential features for lecture-oriented course as well as for a MOOC. Our purpose was to design and build a tool that is functional in both types of courses.
The quantum circuit simulator is implemented as a TIM plugin, and it can be customized to create different types of exercises. The teacher has options to customize the different components to be visible or not according to the context it is used in.

The simulator is usable by different kind of demographics. Using the simulator doesn't require neither physics knowledge nor programming knowledge like Qiskit \cite{Qiskit}. On the other hand, the simulator allows the user to export the circuit to the Qiskit equivalent, or the final quantum state and output probabilities for further analysis if needed.

\subsection{The circuit simulator interface}

\begin{figure}
    \centering
    \includegraphics[width=\columnwidth]{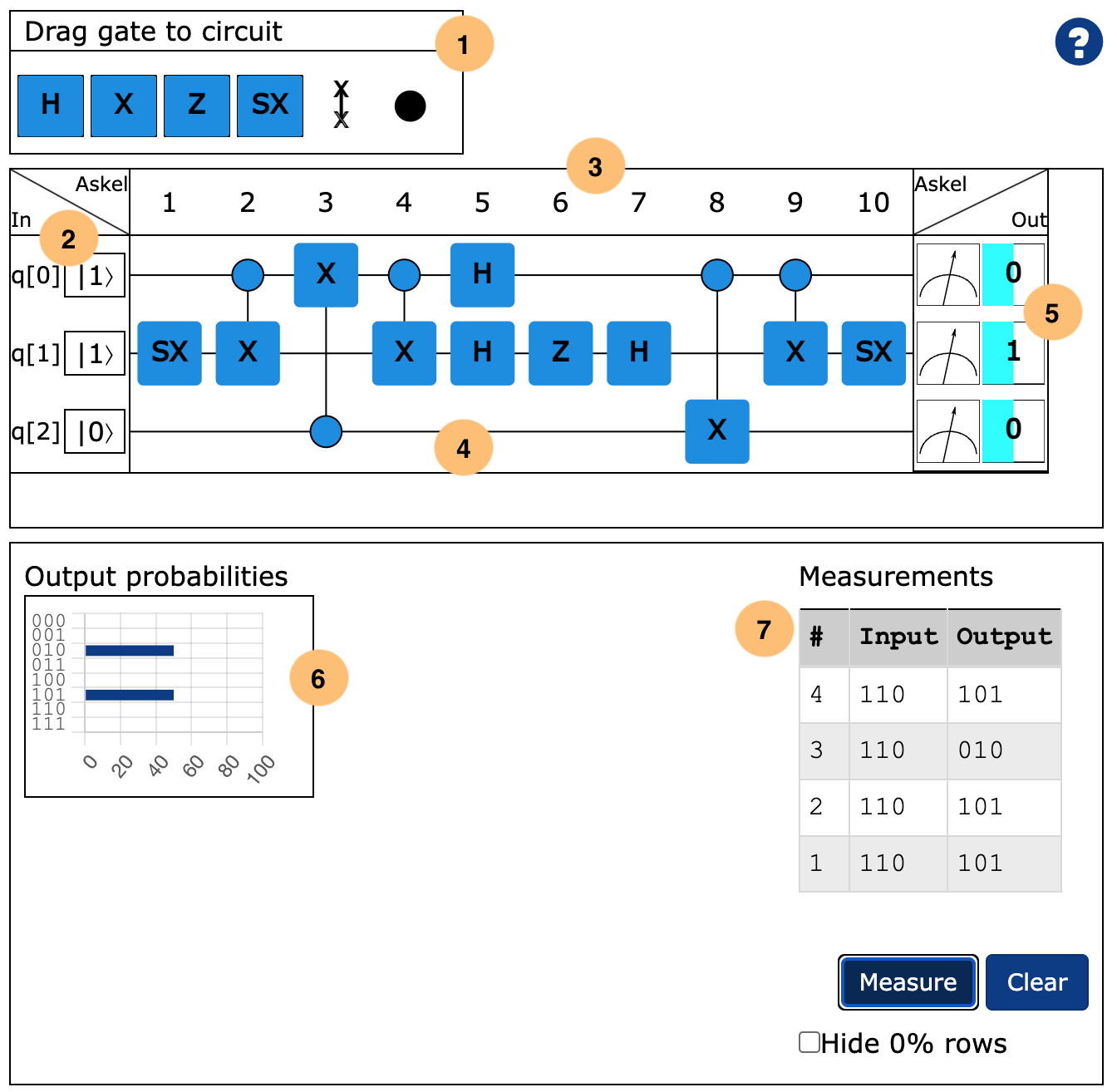}
    \caption{Quantum circuit simulator interface tour: (1) available gates toolbar, (2) input qubits and values, (3) maximum circuit steps, (4) circuit layout grid, (5) qubit output measurements -- value and probability, (6) results probability distribution, and (7) individual execution results table -- inputs and output measurements}
    \label{fig:simulator-ui}
\end{figure}

The inspiration for our quantum circuit simulator was drawn from the existing graphical quantum circuit simulators like IBM Quantum Composer~\cite{ibmcomposer}, Quantum Länd Circuit Simulator~\cite{quantumland}, Quantum JavaScript~\cite{qjs}, Quirck~\cite{quirk} or Uranium~\cite{uranium}. One of the design goals of our circuit simulator was adaptability so that it could be modified to fit different types of exercises \textit{within} the TIM platform. For this reason, we choose to create a new simulator that incorporates the relevant visual elements for designing circuits and visualise the results present in existing simulators.

The circuit simulator contains an \textit{editor} area that allows the user to construct the circuit -- toolbar with available gates, input qubits names and values (e.g. $\ket{0}$ or $\ket{1}$), maximum number of circuit steps, circuit layout grid, and qubit output measurements with probability distribution, and an \textit{results} area for visualizing the simulation results -- probability distribution, or individual results table. The annotated quantum circuit simulator user interface is depicted in Figure~\ref{fig:simulator-ui}.

The editor functionality of the simulator allows the user to construct the circuit from the gates available in the toolbar, by dragging and dropping gates in the circuit layout grid. The controlled gates can be created by dragging control nodes to same step as target gate. The controls are connected by vertical wires to the target gate, while the anti-controls -- e.g., controls that are activated with zero instead of one -- are represented with unfilled circles. The multi-qubit gates are represented as rectangles covering adjacent lines. The default configuration of the simulator contains a set of common gates, which can be extended with own custom gates.

The size of the circuit layout grid is determined by input qubit count for rows and step count for columns, and is immutable. This way user knows exactly how many qubits and steps the answer consists of, which can help to narrow down their choices on how to implemented the desired circuit. Even on simple exercises there are many possible ways to arrange the available gates in the circuit so this predetermined circuit size might make the exercises more manageable for the user.

The user can edit the circuit and see the results in real-time, allowing them to interactively experiment with the behavior of gates to understand how quantum computation works. The progression of the computation is represented as discrete steps going from left to right as gates modify the quantum state. The measurement is done on all qubits at the end in the $\ket{0}$, $\ket{1}$ basis.

The input state of circuit is represented using bits that user can toggle between 0 and 1. being able to change input bit values makes it possible to explore the behavior of the circuit especially when working with simple gates like X and CX. The instructor can decide whether to use braket or bit notation for inputs and give initial values and names for the qubits. The values of the qubits can be locked to specific value if changing the value isn't needed in the exercise.

\begin{figure}
    \centering
    \hspace{5mm}
    \subfloat[\label{circ-a}]{\includegraphics[]{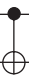}
    }
    \hfill
    \subfloat[\label{circ-b}]{\includegraphics[]{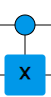}
    }
    \hfill
    \subfloat[\label{circ-c}]{\includegraphics[]{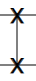}
    }
    \hspace{5mm}
    \caption{Example gate representations in the visual simulator's toolbar: (a) a common notation for a CX-gate, (b) a CX-gate using our unified notation, and (c) a SWAP gate.}
    \label{fig:gates}
\end{figure}

The graphical presentation of the circuit is similar to what is used in the literature \cite{chen201864, chen2022veriqbench}.
For example, the Control-X gate is typically represented as $\oplus$.
Since our simulator allows to add a control dot to any quantum gate, we chose to use more unified notation, see Figure \ref{fig:gates}. 
First, every gate is represented by a rectangle with it's name inside it. A control dot, or several of them, can be dragged and connected to the gate. The only exception to this rule is the swap gate which is represented by two X connected by wire.

The results part of the simulator interface provides several widgets that allow the visualization of the quantum computation. The bar chart widget shows the probability distribution of the output states. For larger circuits, rows where output probability is zero can be hidden from the chart to make the results easier to understand. The measurements widget allows the user to perform shot measurements and to visualize the sampled output value in a table containing the measurement number, and the corresponding input and output.

The simulator computes the full state vector allowing for statistical sampling of the outputs. Samples can be drawn from the theoretical output probability distribution computed from the final state vector. The output probabilities shown in the chart can be these exact probabilities or the probabilities can be based on automatic sampling of certain sample size or from the measurements made by the user.

The visibility of the elements of the circuit simulator interface is configurable, therefore the user can first be shown the basic version and then additional features can be gradually introduced making the simulator more approachable for new users.

\subsection{System architecture}

\begin{figure*}[t!]
    \centering
    \includegraphics[width=0.65\textwidth]{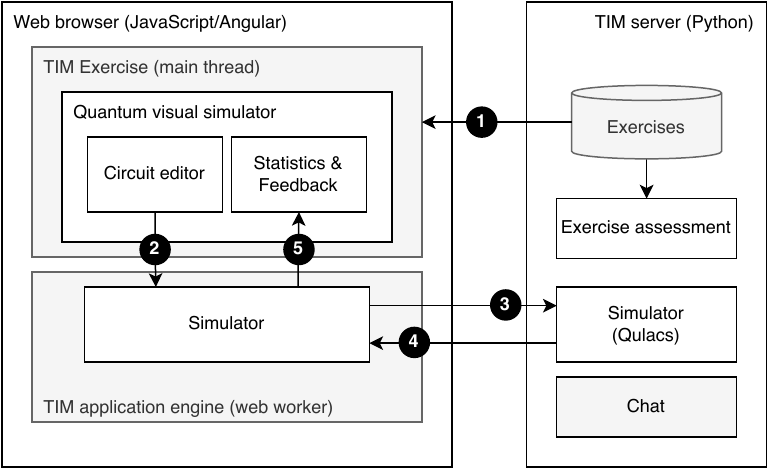}
    \caption{Visual quantum simulator architecture overview (existing TIM components in gray), and interaction between the components: (1) loading the exercise, (2) initialize client-side simulator, (3) initialise server-side simulator if circuit is large, (4) return the results, and (5) render the results}
    \label{fig:arch}
\end{figure*}

The quantum circuit simulator implementation consists of a client component and  server component. The client component, written in JavaScript, using Angular\footnote{\href{https://angular.io}{https://angular.io}} framework, integrates the circuit visualisation into the TIM exercise user interface -- executed into the web browser main thread, and the client-side simulator that integrates into the TIM application engine -- executed in a web worker. The server component implements the server-side circuit simulator, using the Qulacs library~\cite{suzuki2021qulacs}, and the exercise assessment. The interaction between the client-side and server-side simulators is realised via a REST interface. The decomposition of the system is depicted in Figure \ref{fig:arch}.

When opening a new task in the exercise, the user interface renders the circuit editor and initialises the client-side simulator. If the circuit is too large to be handled locally, the server-side simulator is initialized. After initialization, the circuit is executed on the simulator and the results are collected, send to the browser if necessary, then rendered in the user interface. A new execution process is performed whenever the user modifies the circuit or changes the input values, allowing the user interface to be always up to date. To maintain the responsiveness of the user interface, the client-side simulator in implemented in a web worker. Therefore, even under intense load the browser main thread is not blocked.

When there are a lot of simulation exercises on one web page the load time increases making it quite slow especially on slower devices like tablets. This was mitigated by lazy loading the simulator components. The simulator is fully rendered and the simulation computation is done when user clicks on the exercise.

The exercises can be automatically checked for correctness by specifying a model circuit to compare the user's circuit to. The circuit equality is checked by iterating over all $2^{nQubits}$ input combinations and checking that both circuit result in same probability vector. The number of checked inputs can be reduced by giving the regular expressions patterns to match each input bit-strings against, and only running the simulation against these matching bit-string inputs.

\subsection{MOOC integration}

To support the teaching activities, the visual quantum circuit simulator is seamlessly integrated into TIM's MOOC environment. Each task instance is visualized by an instance of the circuit simulator. 
When creating a task, the instructor can control the appearance and behaviour of the simulator using a YAML-based configuration, depicted in Listing \ref{listing:program}. The configuration options can be grouped into the following categories: the task description (lines 1--2), the appearance of the simulator's user interface elements (lines 4--10), the circuit properties -- number of qubits, circuit steps, qubits initial state and editability, initial and target circuit state, and special gate conditions (lines 12--35), and the feedback provided to the student (lines 37--40). The text-based format of the configuration enables the instructor to reuse and adapt circuits for different learning tasks, lowering the effort to create or maintain the course material.

The student attempt to solve a task is saved on the server.  This allows the instructor to asses the progress for every student. The instructor can engage in a conversation with the student using the TIM's chat function, both sharing the same attempt to solve a task context. As the visual simulator is implemented as a TIM plugin, it leverages the interactive capabilities of the MOOC platform to provide the teaching experience that is not possible with individual study tools like IBM Quantum Composer.

\begin{listing}[t]
\begin{minted}[
    fontsize=\scriptsize,
    linenos=true,
    highlightlines={},
    breaklines
]{yaml}
header: "Tehtävä"
stem: "Käytä kahta CX-porttia ja etsi niille sellainen muodostelma, että ensimmäisen bitin arvo q[0] kopioituu kahdelle seuraavalle bitille kun nämä ovat alkutilassa 0. Kun olet saanut piirin valmiiksi, paina 'Tallenna' ja ratkaisusi tarkistetaan."

qubitNotation: "bit"
showChart: false
showOutputBits: true
middleAxisLabel: "Askel"
leftAxisLabel: "In"
rightAxisLabel: "Out"

nQubits: 2
nMoments: 4
gates: ["X","control"]
samplingMode: matrix
qubits:
  - value: 0
    editable: true
  - value: 0
    editable: false
modelCircuit:
  - controls:
      - 0
    editable: true
    name: X
    target: 1
    time: 0
  - controls:
      - 1
    editable: true
    name: X
    target: 2
    time: 1
modelConditions: ["C1X == 2"]
  - pointsRule: # tämän alle säännöt miten pisteitä saa
      multiplier: 5.0 # Millä luvulla kerrotaan tehtävästä saadut pisteet 

feedbackText:
    correct: "Oikein"
    conditionWrong: "Väärin. Vastauksessa pitää olla kaksi CX-porttia."
feedbackShowTable: false
\end{minted}
\caption{YAML representation of a circuit in TIM, with localization properties in Finnish}
\label{listing:program}
\end{listing}

\section{DEMONSTRATION}

The quantum circuit simulator is a flexible tool to construct various exercises.
In the following we demonstrate three different kind of exercises that are used in our online courses.

\subsection{Understanding the probabilistic nature of quantum gates}

\begin{figure}
    \includegraphics[width=\columnwidth]{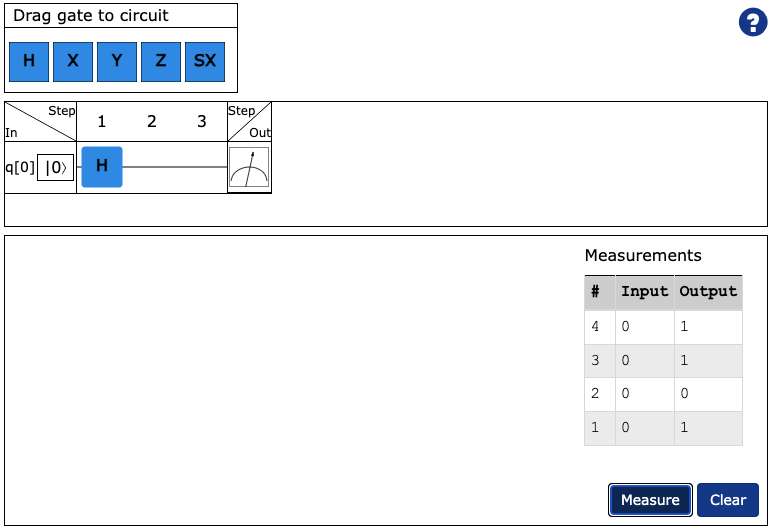}
    \caption{Task 1 - Understanding probabilistic gates}
    \label{fig:task-1}
\end{figure}

Quantum gates, such as Hadamard gate and square-root-X, do not lead to determined measurement outcomes.
They are the simplest examples of non-classical building blocks of quantum circuits and for that reason natural first elements in a quantum computing course.
As one of the first demonstrations of the difference of standard gate based computation and quantum gate computation, a student is given an exercises where she needs to find what gates lead to deterministic outcomes.
In that kind of exercise, the simulator is given with the settings as in Figure \ref{fig:task-1}.
In that setting the circuit gives individual measurement outcomes and the non-deterministic nature of quantum gates is transparent.
Pressing 'measure' several times and getting different outcomes is very concrete and better for this task than a static visualization of a probability distribution.
Furthermore, both Hadamard and square-root-X gates offer intriguing demonstration of quantum randomness as individually used they lead to random outcomes, but two similar gates to deterministic outcomes.

\subsection{Working with quantum circuits}

The action of a single quantum gate is simple. To achieve certain task, one may have to use a circuit composed of several gates. Their collective performance may be hard to infer.
The circuit simulator offers a fantastic workaround as the student is encouraged to experiment with gate constellations. The simulator gives an instant visualization of the output in the form of measurement outcome distribution.

\begin{figure}
    \includegraphics[width=\columnwidth]{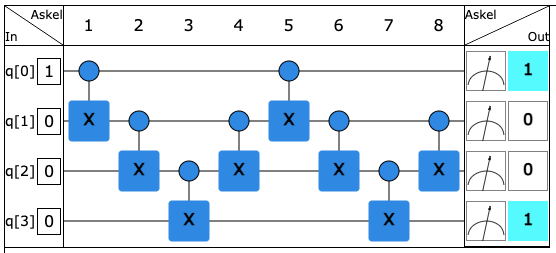}
    \caption{Task 2 - Working with circuits}
    \label{fig:task-2}
\end{figure}

A real quantum computer has certain native gates and all the other gates must be build up from those. For that reason, it is important to learn to recognize circuits with equivalent performance. 
The simulator enables to give complicated gate decompositions (see Figure~\ref{fig:task-2}), and the task for a student is to find equivalent circuit with less gates. In the depicted case one can replace the whole circuit by just one controlled X-gate.
The simulator has an automatic system to check answers, and in this way a student gets an immediate feedback if her trial is the correct one. A typical exercise does not have a unique answer, but the automatic evaluator is capable of recognizing all correct answers from their input-output behaviour.
It is, in fact, common that students find novel solutions that a teacher was not able to foresee.

\subsection{Identifying an unknown quantum gate}

\begin{figure}
    \includegraphics[width=\columnwidth]{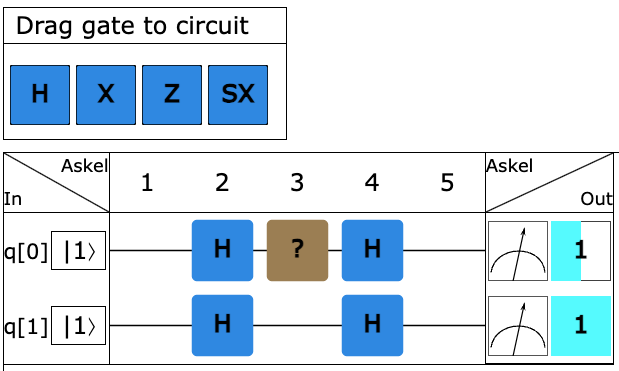}
    \caption{Task 3 - Identifying unknown quantum gate}
    \label{fig:task-3}
\end{figure}

Once a student has reached some familiarity of quantum gates, she can proceed to more interesting exercises. 
With the simulator the teacher can create arbitrary gates and label them in any desired way. This makes it possible to form exercises where a student has to identify an unknown gate, or their combination, by using it in concatenation with other gates (see Figure~\ref{fig:task-3}).
This kind of exercises are almost like entertaining puzzles.
As the gate can be any of the previously introduced gates, a student necessarily has to review previously learned material.
Some quantum gates, like the Z-gate, do nothing in the computational basis unless they are coupled with some other, appropriately chosen, quantum gates.
This quantum feature makes additional excitement to suitably planned exercises.

\section{DISCUSSION}

\begin{figure*}
    \centering
    \includegraphics[width=0.65\textwidth]{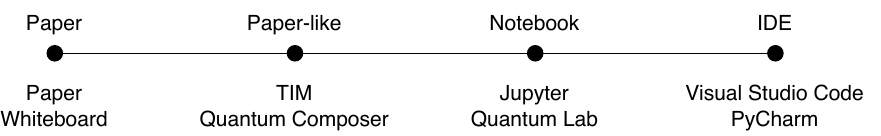}
    \caption{Spectrum of tools that can be used by quantum information and computing practitioners}
    \label{fig:spectrum}
\end{figure*}

\subsection{Positioning among development tools}

The spectrum of experiences that can be used by quantum technology practitioners (excluding hardware), can be divided into the following areas, as depicted in Figure \ref{fig:spectrum}: paper, paper-like, notebook, and integrated development environment (IDE). The paper is used mainly by practitioners with physics and mathematical background and does not necessitate computer systems for experimenting besides paper or a whiteboard. The paper-like provides a low threshold for using a computer system that offers an experience close to the paper, but with instant feedback aided by tools like TIM or IBM Quantum Composer. The notebook experience is familiar to scientist that are using Python programming language and quantum development toolkits (e.g., Qiskit, Cirq\footnote{\href{https://quantumai.google/cirq}{https://quantumai.google/cirq}} or Pennylane\footnote{\href{https://pennylane.ai}{https://pennylane.ai}}), and use tools like Jupyter and IBM Quantum Lab to develop algorithms and perform experiments. The IDE experience is tailored for software developers that use advanced software development tools like Visual Studio Code\footnote{\href{https://code.visualstudio.com}{https://code.visualstudio.com}} or PyCharm\footnote{\href{https://www.jetbrains.com/pycharm/}{https://www.jetbrains.com/pycharm/}}.

As the visual appearance of the TIM circuit simulator was inspired by the IBM Quantum Composer, we can say that their core functionality is similar, allowing the users to design and execute quantum circuits using metaphors specific to graphic user interface, and are able to explore their essential characteristics with specialised widgets. 
However, as the purpose of each tool varies -- TIM is a tool optimised to facilitate learning in a formal education environment, while Quantum Composer is a commercial tool designed to smooth the onboard process of developers into IBM's quantum platform -- we expect the latter to add new visual programming capabilities and integrations with advanced simulators and actual quantum hardware at a much higher pace. Nevertheless, by being familiarized with the same metaphors and core capabilities, TIM's users will be able to easy transition to more advanced commercial tools.

\subsection{User feedback}

\begin{figure}
    \includegraphics[width=\columnwidth]{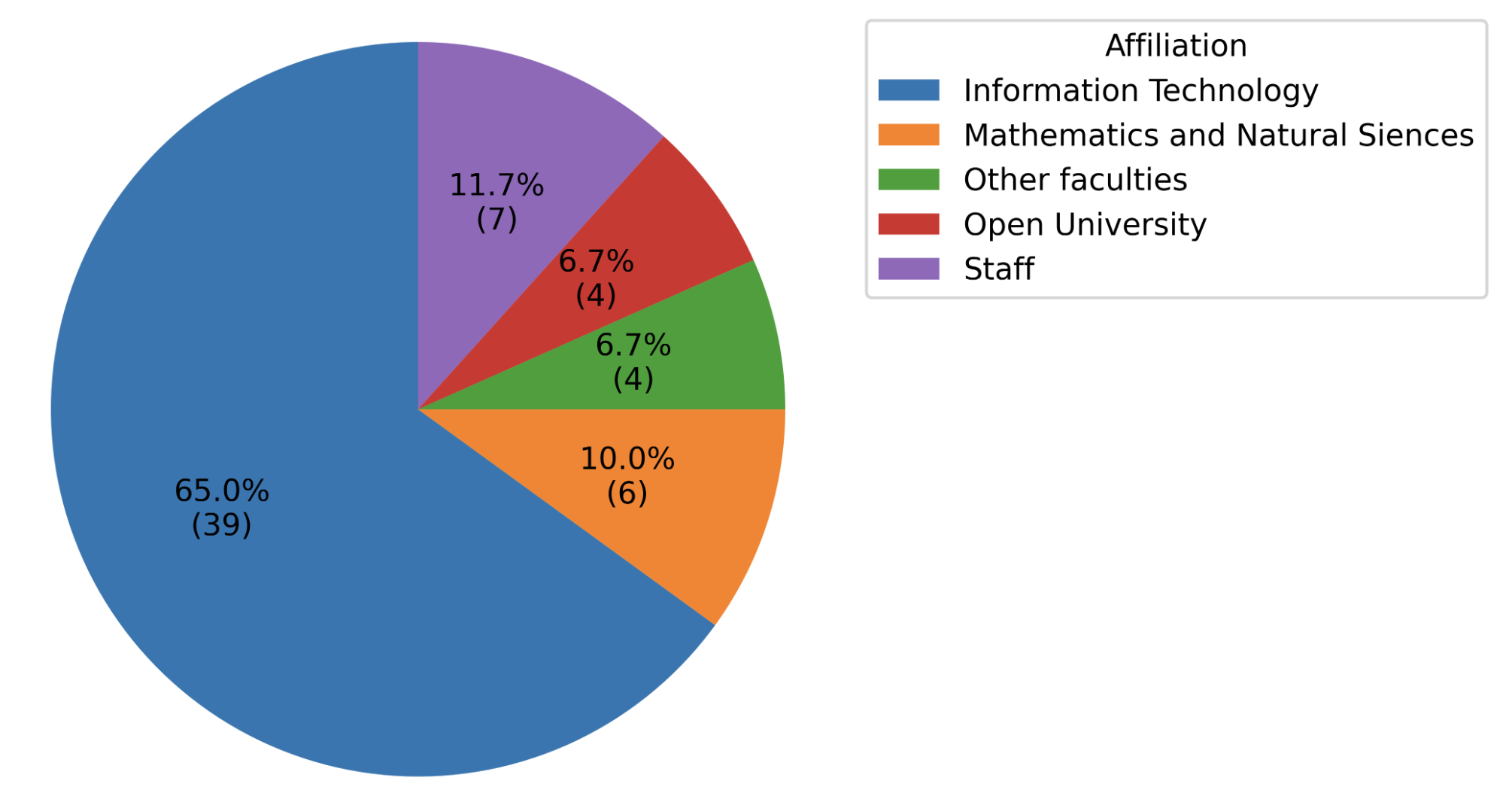}
    \caption{Participants breakdown by department affiliation}
    \label{fig:feedback}
\end{figure}

Since February 2024, we have been running the first MOOC for a group of 60 early adopters (see the details in Figure \ref{fig:feedback}). The Quantum Computing Essentials course is the first in a series of three courses planned at University of Jyväskylä. The language used is Finnish, and participants receive 2 study credits upon completion. The largest group is formed of students from the Information Technology, and Mathematics and Natural Sciences faculties that study quantum computing as part of their curriculum. Another large segment was formed by the university staff, followed by students from faculties that do not mandate the study of quantum computing, and participants from the open university. The study group covers two from the three student personas described earlier (e.g. physicist and software engineer), allowing us to draw some preliminary observations.

Given the diverse backgrounds of the students for whom the MOOC is designed, we anticipated that the difficulty of tasks would vary among students. 
In order to prevent any task from becoming a significant bottleneck, we included additional hints in the course material that effectively served their purpose.
To measure the difficulty of a task, we have used the average number of attempts made by the student before reaching the correct answer. Since the number of attempts does not impact the course grade, students are motivated by the opportunity to tackle the task challenges without the pressure of exam failure or low grades. 
In this regard, the MOOC format is functioning as intended, fostering a supportive environment for learning.

We recall that the simulator’s primary advantage over similar products is its seamless integration with the existing TIM learning platform. It’s not confined to a single course but can be utilized across various courses, allowing for the easy generation of exercises with diverse difficulty levels. The integration with TIM provides access to all necessary tools for a university course, including discussion forums, straightforward registration, and grading.
The value of these features has been demonstrated, and the simulator has fulfilled its intended purpose.
Among our target student groups, the paper-like experience provided by the TIM's quantum circuit simulator offers the optimal experience, as it does not require familiarity neither with programming languages, nor mathematical concepts, instead leading students to experiment with circuits by drag-and-dropping gates and instantly see the results.

Very recently, the MOOC was started in the open university, where it is offered for free. This will be a real-world test of our primary objective: making quantum computing accessible to everyone.

\def\refname{REFERENCES}

\bibliographystyle{plain}

\begin{thebibliography}{10}

\bibitem{ibmcomposer}
Ibm quantum composer, 2024.
\newblock Last accessed 22 February 2024. https://quantum.ibm.com/composer/files/new.

\bibitem{quantumland}
Quantum circuit simulator - the quantum länd, 2024.
\newblock Last accessed 22 February 2024. https://thequantumlaend.de/quantum-circuit-designer/.

\bibitem{qjs}
Quantum javascript (q.js), 2024.
\newblock Last accessed 22 February 2024. https://github.com/stewdio/q.js.

\bibitem{quirk}
Quirk: Quantum circuit simulator, 2024.
\newblock Last accessed 22 February 2024. https://algassert.com/quirk.

\bibitem{uranium}
Uranium circuit-editor, 2024.
\newblock Last accessed 22 February 2024. https://uranium.transilvania-quantum.org/circuit-editor/.

\bibitem{chen2022veriqbench}
Kean Chen, Wang Fang, Ji~Guan, Xin Hong, Mingyu Huang, Junyi Liu, Qisheng Wang, and Mingsheng Ying.
\newblock Veriqbench: A benchmark for multiple types of quantum circuits.
\newblock {\em arXiv preprint arXiv:2206.10880}, 2022.

\bibitem{chen201864}
Zhao-Yun Chen, Qi~Zhou, Cheng Xue, Xia Yang, Guang-Can Guo, and Guo-Ping Guo.
\newblock 64-qubit quantum circuit simulation.
\newblock {\em Science Bulletin}, 63(15):964--971, 2018.

\bibitem{TIMLessIsMore2019}
Ville Isom{\"o}tt{\"o}nen, Antti-Jussi Lakanen, and Vesa Lappalainen.
\newblock Less is more! preliminary evaluation of multi-functional document-based online learning environment.
\newblock In {\em 2019 IEEE Frontiers in Education Conference (FIE)}, pages 1--5, 2019.

\bibitem{doi:10.2753/MIS0742-1222240302}
Marcus A.~Rothenberger Ken~Peffers, Tuure~Tuunanen and Samir Chatterjee.
\newblock A design science research methodology for information systems research.
\newblock {\em Journal of Management Information Systems}, 24(3):45--77, 2007.

\bibitem{Qiskit}
{Qiskit contributors}.
\newblock Qiskit: An open-source framework for quantum computing, 2023.

\bibitem{suzuki2021qulacs}
Yasunari Suzuki, Yoshiaki Kawase, Yuya Masumura, Yuria Hiraga, Masahiro Nakadai, Jiabao Chen, Ken~M Nakanishi, Kosuke Mitarai, Ryosuke Imai, Shiro Tamiya, et~al.
\newblock Qulacs: a fast and versatile quantum circuit simulator for research purpose.
\newblock {\em Quantum}, 5:559, 2021.

\end{thebibliography}

\begin{IEEEbiography}{Juha Reinikainen}{\,} is a master's degree student in mathematical information technology in the University of Jyväskylä. His research interests include quantum computing, simulation and software development. Contact him at juha.a.reinikainen@jyu.fi.
\end{IEEEbiography}

\begin{IEEEbiography}{Vlad Stirbu}{\,} is a postdoctoral researcher at the University of Jyväskylä, Finland. His research interests include: software engineering, software architecture, quantum software, and software development in regulated industries. He is a member of IEEE Computer Society. Contact him at vlad.a.stirbu@jyu.fi.
\end{IEEEbiography}

\begin{IEEEbiography}{Teiko Heinosaari}{\,} is a professor in quantum computing at the University of Jyväskylä. His research encompasses various topics in quantum information theory and hybrid classical-quantum computing. He is developing novel ways to teach quantum computing to students who have no prior knowledge of quantum physics. Contact him at teiko.heinosaari@jyu.fi.\vadjust{\vfill\pagebreak}
\end{IEEEbiography}

\begin{IEEEbiography}{Vesa Lappalainen}{\,} is a senior lecturer at the University of Jyväskylä with over 40 years of teaching experience. He is the creator of the TIM-platform. His current research interest is e-learning. Contact him at vesa.t.lappalainen@jyu.fi.
\end{IEEEbiography}

\begin{IEEEbiography}{Tommi Mikkonen}{\,} is a professor of software engineering at the University of Jyväskylä, in Finland. Contact him at tommi.j.mikkonen@jyu.fi.
\end{IEEEbiography}

\end{document}